\documentclass{article}

\usepackage{arxiv}

\usepackage[utf8]{inputenc} 
\usepackage[T1]{fontenc}    
\usepackage{hyperref}       
\usepackage{url}            
\usepackage{booktabs}       
\usepackage{amsfonts}       
\usepackage{nicefrac}       
\usepackage{microtype}      
\usepackage{lipsum}
\usepackage{amssymb}
\usepackage{latexsym}
\usepackage{amssymb}
\usepackage{epsfig}
\usepackage{graphicx}
\usepackage{amsmath}
\usepackage{amssymb}
\usepackage{algorithm}
\usepackage{algorithmic}
\usepackage{subfigure}
\usepackage{caption}
\usepackage{epsfig}
\usepackage{epstopdf}
\usepackage{cite}
\usepackage{multirow}
\usepackage{url}
\usepackage{xcolor}

\usepackage{hyperref}

\title{Whole Slide Images based Cancer Survival Prediction using Attention Guided Deep Multiple Instance Learning Networks\thanks{Preprint, to appear in Medical Image Analysis 65, 101789, 2020 (\url{https://doi.org/10.1016/j.media.2020.101789})}}

\author{
  Jiawen Yao \\
  Department of Computer Science and Engineering\\
  University of Texas at Arlington\\
  Arlington, Texas, USA \\
  \texttt{yjiaweneecs@gmail.com}\\
   \And
 Xinliang Zhu \\
  Department of Computer Science and Engineering\\
  University of Texas at Arlington\\
  Arlington, Texas, USA \\
     \And
 Jitendra Jonnagaddala \\
  School of Public Health and Community Medicine\\
  University of New South Wales\\
  Sydney, NSW, Australia \\
     \And
 Nicholas Hawkins \\
  School of Medical Sciences\\
  UNSW Sydney\\
  Sydney, NSW, Australia \\
     \And
 Junzhou Huang\thanks{Corresponding author} \\
  Department of Computer Science and Engineering\\
  University of Texas at Arlington\\
  Arlington, Texas, USA \\
  \texttt{jzhuang@uta.edu} \\
}

\begin{document}
\maketitle

\begin{abstract}
Traditional image-based survival prediction models rely on discriminative patch labeling which make those methods not scalable to extend to large datasets.  Recent studies have shown Multiple Instance Learning (MIL) framework is useful for histopathological images when no annotations are available in classification task. Different to the current image-based survival models that limit to key patches or clusters derived from Whole Slide Images (WSIs), we propose Deep Attention Multiple Instance Survival Learning (DeepAttnMISL) by introducing both siamese MI-FCN and attention-based MIL pooling to efficiently learn imaging features from the WSI and then aggregate WSI-level information to patient-level. Attention-based aggregation is more flexible and adaptive than aggregation techniques in recent survival models. We evaluated our methods on two large cancer whole slide images datasets and our results suggest that the proposed approach is more effective and suitable for large datasets and has better interpretability in locating important patterns and features that contribute to accurate cancer survival predictions. The proposed framework can also be used to assess individual patient's risk and thus assisting in delivering personalized medicine. Codes are available at \url{https://github.com/uta-smile/DeepAttnMISL_MEDIA}

\end{abstract}

\keywords{Survival Prediction\and Multiple Instance Learning\and Deep Learning\and Whole Slide Images}

\section{Introduction}
\noindent Survival analysis aims to analyze the expected duration of time until events happen. It tries to find the answer of questions like: how does the proportion of a population survive past a certain time (e.g. 5 years)? What rate will they die or fail?
It is a very important clinical application and many efforts have been made to search for biomarkers from omics data that are significantly related to patient death \cite{shedden2008gene,tibshirani1997lasso,bair2004semi,bair2006prediction,park2007l1}. Recent technological innovations are enabling scientists to capture big whole slide images (WSIs) at increasing speed and resolution for diagnosis. The learning model is required to correctly predict the survival risk of each patient from his/her tumor tissue whole slide images. The more precise is risk assessment for a cancer patient, the better the patient can be treated. Compare with genomics data, pathological images can present tumor growth and morphology in extremely detailed, gigapixel resolution which is extremely useful for cancer study~\cite{warth2012novel, yuan2012quantitative}.

The diagnosis is extremely laborious and highly dependent on expertise which requires pathologists to carefully examine the biopsies under the microscope~\cite{bejnordi2017diagnostic}. To reduce the risk of misdiagnosis, pathologists have to conduct a thorough inspection of the whole slide which make the diagnosis quite cumbersome. Automatic analysis of histology has become one of the most rapidly expanding fields in medical imaging. Computer aided diagnostics in digital pathology can not only alleviate pathologists' workloads, but also help to reduce the chance of diagnosis mistakes. However, using WSIs for survival prediction is very challenging due to several reasons: 1) pathological images in real cancer dataset might be in terabytes ($10^{12}$ pixels) level which makes most models computationally impossible. 2) the large variations of textures and biological structures from tumor heterogeneity, As the solid tumor may have a mixture of tissue architectures and structures, multiple WSIs from different parts of the patient's tissue are collected for diagnosis;  3) label on patient-level while each patient might have multiple WSIs for diagnosis. Those terabyte-size large WSIs from one patient will share the survival label which will make the problem more challenging.



\subsection{Related Work}
During recent years, many methods have been proposed for survival prediction using pathological slides. They can be categorized into two categories: ROI-based and WSI-based methods.

\textbf{Region of Interest Analysis.}
Pathological images usually come with a very high resolution which makes most of exiting models and algorithms computationally infeasible even though the high resolution of image data greatly benefits survival analysis with more precise information.
Previously due to the lack of computational power, most of the literature focused on regions of interest (ROI) patches which are selected by pathologists from WSIs~\cite{gurcan2009histopathological}. 

Instead of handling original WSIs, ROI-based methods extracted hand-crafted features from ROIs for predictions~\cite{yuan2012quantitative,Barker201660,Zhu2016Lung,yao2016imaging,Wang2014Novel,yu2016predicting, cheng2017identification,yao2015computer}. Wang et al.~\cite{Wang2014Novel} proposed a novel framework to first segment cells in annotated patches and then perform cellular morphological properties from those cells which result in 166 imaging features. Yu et al.~\cite{yu2016predicting} extract 9,879 quantitative image features from annotated regions of interest and results suggest that automatically derived image features can predict the prognosis of lung cancer patients and thereby contribute to precision oncology. Beyond classical cell detection, Yao et al.~\cite{yao2016imaging} used a deep subtype cell detection first to classify different cell subtypes and then extracted features from cellular subtype information. Cheng et al.~\cite{cheng2017identification} used a deep auto-encoder to cluster cell patches into different types and then extracted topological features to characterize cell type distributions from ROIs for prediction. These methods extracted hand-crafted features based on nuclei detection and segmentation and those features were considered to represent prior knowledge of boundary, region or shape. However, hand-crafted features are limited in representation power and capability.

Recently, with the advance of deep neural networks,  deep learning-based survival models are proposed for seeking more powerful deep representation~\cite{Katzman2016deepsurv, zhu2016deep, yao2017deep, mobadersany2018predicting}. 
Katzman $et~al.$ first proposed a deep fully connected network (DeepSurv) to represent the nonlinear risk function~\cite{Katzman2016deepsurv}. They demonstrated that DeepSurv outperformed the standard linear Cox proportional hazard model. Another improvement is deep convolutional survival learning (DeepConvSurv) which is the first attempt to use pathological images in deep survival model~\cite{zhu2016deep}. Later, Yao et al.~\cite{yao2017deep} integrated genome modality with DeepConvSurv for survival prediction using multi-modality data. However, DeepConvSurv is designed to use pre-selected ROI patches by pathologists from WSIs for convolution operations. A small set of image tiles might not completely and properly reflect the patients' tumor morphology. Also, 
those methods perform average pooling to achieve patient-wise predictions from patch-based results. Such combination cannot effectively aggregate predictions from patch-level and needs further attention.
Thus, it would be much helpful if we can facilitate knowledge discovery from big whole slide images.

\textbf{Whole-slide Image Analysis.}
With detailed and densely annotations on WSIs, nowadays a series of approaches in whole-slide image analysis have been proposed for a variety of applications including classification, detection or segmentation~\cite{wang2016deep, bejnordi2017diagnostic, kong2017cancer, li2018cancer, liu2017detecting}. Applying deep learning for supervised learning on computational pathology has achieved promising results. However, the applicability of these models in clinical practice remains in questions because of the wide variance of clinical samples. Extensive and time-consuming human manual annotations in clinical practice is impossible. Moreover, the success of those applications is built on integrating detailed patch contents and using labor-extensive annotations which might not be applicable for survival prediction.

To properly address the shortcomings of current models, one possible direction is to consider weakly supervised manner. Recently, researchers have developed many weakly supervised algorithms to medical images including weakly-supervised X-rays screening~\cite{wangyi2019weakly, yan2018weakly} and WSI classification~\cite{Hou2016PatchbasedCN, mercan2018multi, wang2018weakly, wang2019rmdl, wang2019weakly}.
WSI classification models are designed to find the most differentiated regions correspond to different tumor types. A two-step approach is usually used and the first step is a classifier at the tile level and then predicted scores for each tile within a WSI are aggregated with various strategies. However, learning survival from histology and developing prognosis model is considerably more difficult as risk is often reflected from a range of histology patterns that correspond to varying degrees of disease progression. Tumor heterogeneity plays an important role in cancer study which includes inter-tumor and intra-tumor heterogeneity~\cite{jamal2015translational}. Inter-tumor heterogeneity refers to the differences found between tumors in different patients. Intra-tumor heterogeneity refers to distinct tumor cell populations within the same tumor specimen. Most recent weakly-supervised WSI classification focused on localizing most differentiated regions correspond to tumor types across patients. Therefore, they are more likely to capture inter-tumor heterogeneity between tumors or subtype tumors. Understanding how to label one person's tumor type may not be enough to study the degree of tumor progression. The pathophysiology of tumor progression and proliferation is complex 
and thus a new image-based prognosis model which can integrate information from heterogeneous tissue regions is a better approach. 
Additionally, existing weakly-supervised WSI classification task is at slide-level, while survival prediction is at patient-level analysis (one patient might have multiple whole slide images). Devising patient-level decisions from slide-level results is not the objective of those studies.
To achieve survival prediction from whole slide images without using annotations, Zhu et al.~\cite{zhu2017wsisa} proposed a patch-based two-stage framework to predict patients' survival outcomes.  Patches are extracted from the WSIs and clustered to different patterns defined as "phenotypes" according to their visual appearances in the first stage. 
Then WSISA~\cite{zhu2017wsisa} adopted DeepConvSurv~\cite{zhu2016deep} to select important patch clusters and then aggregated those clusters for final prediction.  Although this framework has practical merits to consider important patch clusters, it is hard to incorporate it into state-of-the-art deep learning paradigm as the whole approach has separate steps. In addition, it is not a scalable solution because the first stage will be significantly inefficient if more patches are sampled. One recent work ~\cite{tang2019capsurv} proposed CapSurv by introducing Capsule network~\cite{sabour2017dynamic}. However, CapSurv still has similar issues with WSISA as the main framework is following the WSISA pipeline. 
The relationship of tissue patterns on WSI is the great importance on survival analysis. Li et al.~\cite{li2018graph} proposed a graph convolutional network (GCN) based method to consider such relationship of patches in the WSI and then learn effective representation for survival prediction. However, this method requires detailed graph structure knowledge to construct a complete graph representation for effective GCN training which is not flexible and needs prior knowledge.

\subsection{Contributions}

Though many works can be found on WSI analysis for segmentation, classification and detection, there were limited works on weakly-supervised learning for survival prediction. Based on the literature review, a method that can adaptively learn patient-level representations with limited prior knowledge is needed. In this study, we propose a novel framework, referred to as Deep Attention Multiple-Instance Survival Learning (DeepAttnMISL) for whole slide images. In contrast to the standard supervised learning, multiple instance learning (MIL) considers a set of bags, each containing multiple feature vectors referred to as instances. The available label is only assigned to bag-level and labels of individual instances in the bag are not known. In MIL, not all the instances are necessarily relevant and some of them in the bag might not be relevant to certain labels. In observation, if the slide is from a low risk patient, most of its tiles might be benign and or contain low-grade tumor. In contrast, if the slide is from the high risk patient, it must be true that at least one of all of the possible tiles contains malignant tumor. This formalization of the WSI survival learning problem is an example of the general standard multiple instance assumption and thus MIL is a good to fit to solve such problem.

Our preliminary work that only using deep multiple instance learning can help achieve better prognosis performance was published in MICCAI 2019~\cite{yao2019deep}. Compared to the previous work, we offered new contributions in following aspects. We introduced attention mechanism into deep multiple instance survival learning. The proposed DeepAttnMISL not only uses the siamese MI-FCN network to learn features from different phenotype clusters, but also largely improves performance with Attention-based MIL pooling layer to perform a trainable weighted aggregation.
More importantly, the proposed framework can effectively highlight the prognosis-related clusters and has better interpretability as well as performance than our preliminary work~\cite{yao2019deep}. The contributions can be summarized as follows.


\begin{itemize}
    \item Phenotype clusters provide morphology-specific representation, the proposed DeepAttnMISL first extracts phenotype-level information through a Siamese MIL-based network from patch-level features. The attention mechanism is then used to aggregate these phenotype features into patient-level information with a trainable weighted average where weights can be fully parameterized by neural networks. Such attention-based aggregation is much flexible than fixed pooling operators in recent work~\cite{yao2019deep, zhu2017wsisa,tang2019capsurv}.
    \item With the advantage of MIL and attention mechanism, the proposed model has a good interpretability to find important patterns of patients. Those identified important regions and patches are more likely to be associated with prognosis and overall the proposed model can achieve better patient-level predictions and improve prediction performance than our previous work~\cite{yao2019deep}.
    
    \item To evaluate the performance of the proposed DeepAttnMISL model, two large WSI datasets on lung and colorectal cancer are used and extensive experimental results verify the effectiveness.

\end{itemize}

Our method can efficiently exploit and utilize all discriminative patterns in whole slide pathological images to perform accurate patients' survival predictions. Additionally, we present results representing a patient's treatment group to illustrate how to view the proposed model as a treatment recommender system. Results validate that the proposed model can accurately model the risk functions of the population and thus guide treatment decisions for improving patient lifespan.

\begin{figure*}[htb]
	\centering
	\includegraphics[width=1\linewidth]{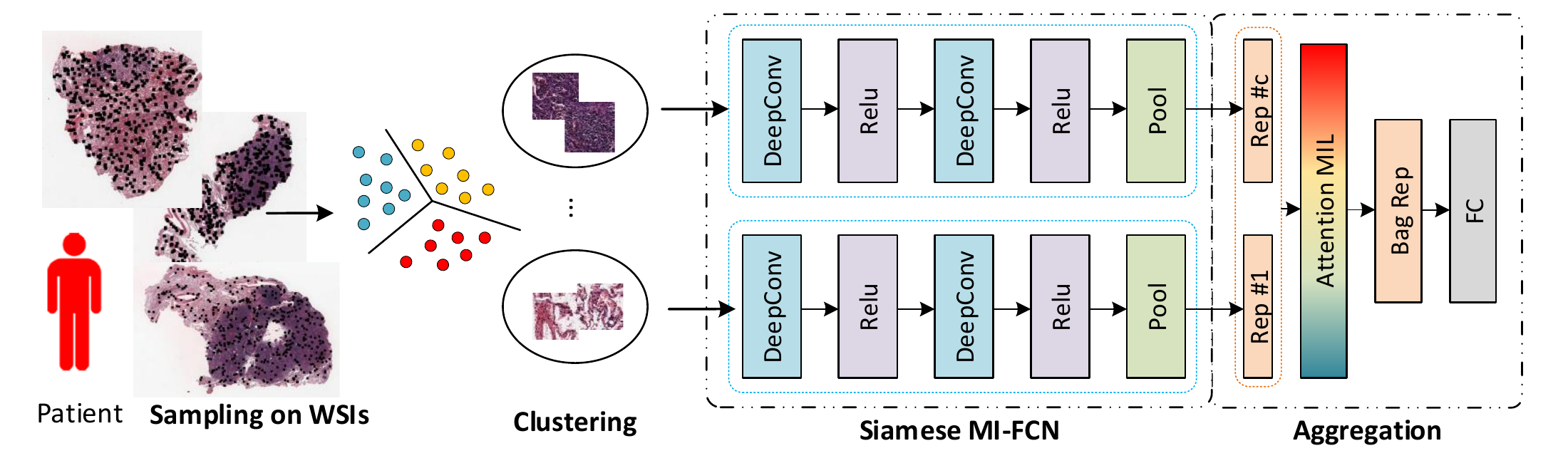}
	\caption{An overview of the proposed DeepAttnMISL model.}
	\label{fig: Framework}
\end{figure*}

\section{Methodology}

Considering a set of $N$ patients, $\{X_i\}, i=1 \dots N$, each patient has the follow-up label $(t_i, \delta_i)$ indicating the overall survival. The observation time $t_i$ is either a survival time or a censored time for each patient.
$\delta_i$ is the indicator which is 1 for an uncensored instance (death occurs during the study) and 0 for a censored instance.
Survival model predicts a value of a target variable O for a given patient. As we discussed above, patient $X_i$ will have multiple WSIs and our goal is to predict the corresponding target $o_i$ from those imaging data. As we don't have pixel-level annotations but only know patient-level information, this weakly-supervised learning can be solved by Multiple Instance Learning (MIL).

In the case of MIL problem, patient $X$ is a bag of instances, $X = \{x_1, ..., x_C\}$ and the number of instances $C$ could vary for different bags. Furthermore, we assume that individual true labels exist for the instances within a bag, i.e., $y_1, ..., y_C$ but those values remain unknown during training. One very important assumption is that neither ordering nor dependency of instances within a bag and a MIL model must be permutation-invariant.
Instances within the bag can be defined as sampling patches from WSIs and several studies~\cite{campanella2019clinical,Wulczyn2020} developed MIL-based deep learning approaches for automated cancer diagnosis and prognosis. In our case, we introduce phenotype cluster as the instance of the bag instead of individual patch. Cancer histology contains rich phenotypic information that reflects underlying molecular processes and disease progression. Phenotype of the pathological slides is a combination of tissue's various observable characteristics. This provides a convenient visual representation of disease aggressiveness. Recent studies have shown phenotypic information could be useful for prediction of prognosis~\cite{zhu2017wsisa,mobadersany2018predicting}. The purpose of the proposed framework is to predict patient outcomes from whole slides images. The study involves partitioning the original slides into a number of phenotype patterns. Each phenotype describes a type of histology pattern and includes a number of smaller patches or tiles.

\subsection{DeepAttnMISL}

Fig.\ref{fig: Framework} shows the overview of the proposed Deep Attention Multiple Instance Survival Learning (DeepAttnMISL). In Multiple Instance Learning, each data sample is a bag of instances and the bag can be seen as one patient in our problem. Each patient $X_i$ may contain multiple whole slides and it is not practical to use whole slides as instances due to the extreme large size. We choose phenotypes instead of raw sampling patches as instances within the bag because it will considerably reduce the complexity of the problem as the number of heterogeneous patches is actually very huge. By using phenotype patterns which are constructed by clustering, we can build the model for different types of tissues to extract morphology-specific features. To learn patient-level information from phenotype clusters, we design a Multiple Instance Fully Convolutional Network (MI-FCN) running inside our deep learning architecture with weights being shared among them as in the siamese architecture. To detect important phenotypes associated with patients' clinical outcomes, attention-based MIL pooling layer is used to aggregate phenotype-level representation. The output is the hazard risk to represent how well for the patient behaves in the population of certain type of diseases.

\subsubsection{Sampling and Clustering}
\begin{figure}[!htb]
	\centering
	\includegraphics[width=0.7\linewidth]{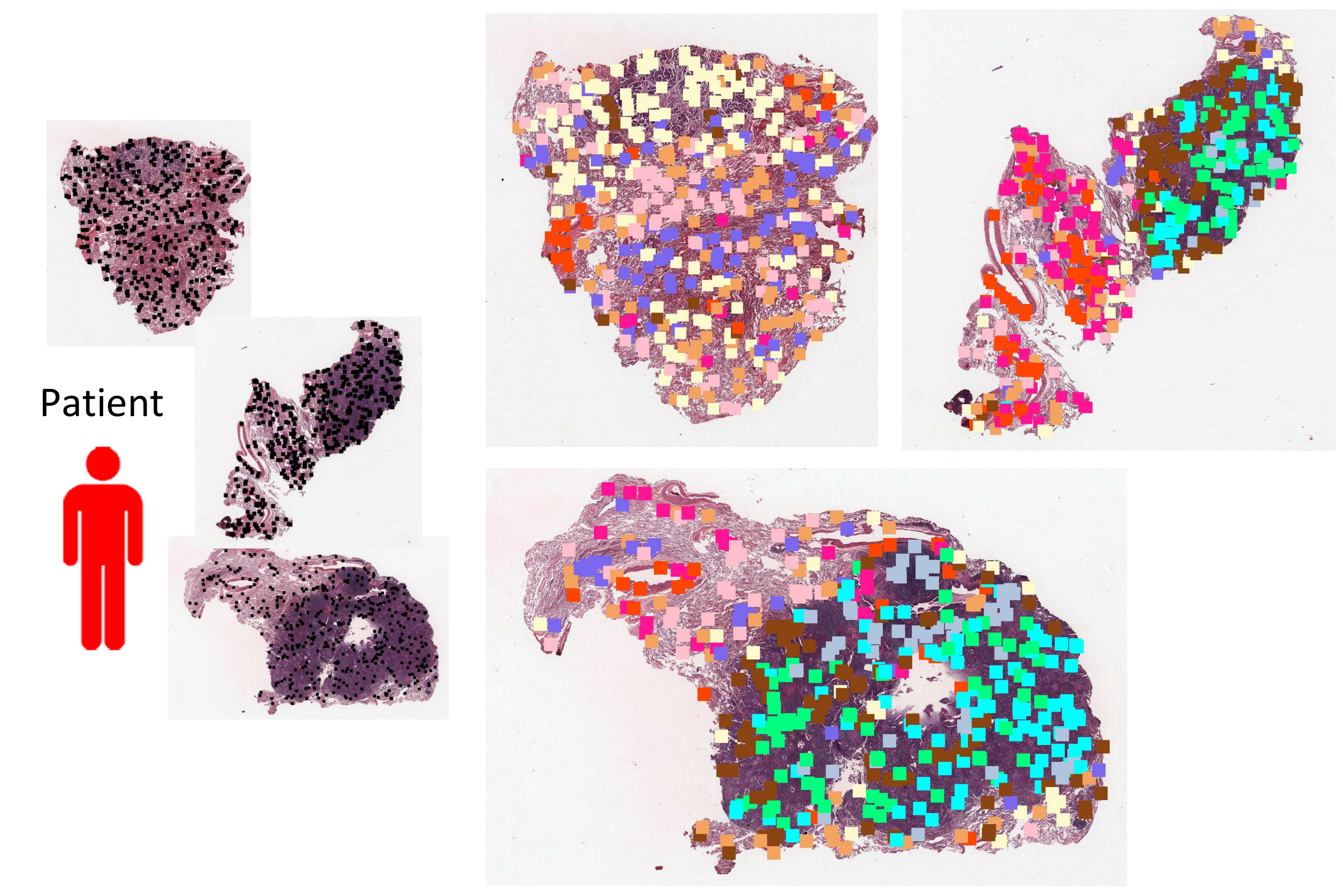}
	\caption{Phenotype patterns visualization after clustering on three WSIs belong to the same patient.}		
	\label{fig:clustering}
\end{figure}

At the first step, we extract patches from all WSIs belong to the same patient and then cluster them into different phenotypes. To capture detailed information of the images, those patches are extracted from 20X (0.5 microns per pixel) objective magnifications and then fixed to $500 \times 500 \times 3$ size. In one whole slide image, usually about 50\% of areas are background and it is easy to select regions to contain tissues rather than background or irregular regions according to pixel values. Even we only extract tissue patches and ignore background regions, it can still get tens of thousands of patches per WSI which will result in a huge number of images from the whole dataset. Different from recent segmentation and detection task in whole slide image analysis, our task is for patient-level decision aggregated from patch-level results. As pointed out in~\cite{hou2015efficient}, training patch-based CNNs for weakly supervised learning is very time costly (several weeks) and we propose to use features from pre-trained models instead of using CNNs to learn features from the scratch. We use the pre-trained model (e.g. VGG) from ImageNet~\cite{simonyan2014very} to extract features for each image patch which have more representation power than smaller size (50 $\times$ 50) thumbnail images to represent their phenotypes~\cite{zhu2017wsisa}. Then we adopt K-means clustering to cluster patches based on their deep learning features. Notice that one patient might have multiple WSIs and we actually perform clustering on patient-level instead of the whole database. Fig.\ref{fig:clustering} shows one patient's example and this patient has three WSIs that were sampled from different locations of the biopsy tissue. The corresponding phenotype clustering are shown in the right and each color means one type of phenotype clusters. In this example, we chose to cluster 10 phenotype patterns. The results show the effectiveness of this strategy as we can see similar patches are grouped into the same cluster.  This could demonstrate that features from pre-trained model are capable of identifying patterns of whole slide images and we would expect them to be distinctive and informative for later survival learning task. 

By clustering different patches from all WSIs of the patient into several distinguished phenotype groups, we will have different phenotype groups with various prediction powers on this patient’s clinical outcome. The proposed DeepAttnMISL takes phenotypes as multiple inputs and consider their connections for predicting survival outcomes.

\subsubsection{Siamese MI-FCN}

After clustering, the patient is a set of phenotype clusters and we design a siamese Multiple Instance Fully Convolutional
Networks (MI-FCN) to learn features from those patterns, \textcolor{red}{similar to the work in \cite{yao2019deep}}. Most existing well-known pre-trained models were trained based on single-instance bases, and the labels are associated with each image which is not the case of our problem. We embed multiple sub-networks running inside our deep learning architecture with weights being shared among them as in the siamese architecture. Each sub-network is based on fully convolutional neural networks (FCN) that can learn informative representation for individual phenotype of the patient. 

\begin{figure}[!htb]
	\centering
	\includegraphics[width=0.6\linewidth]{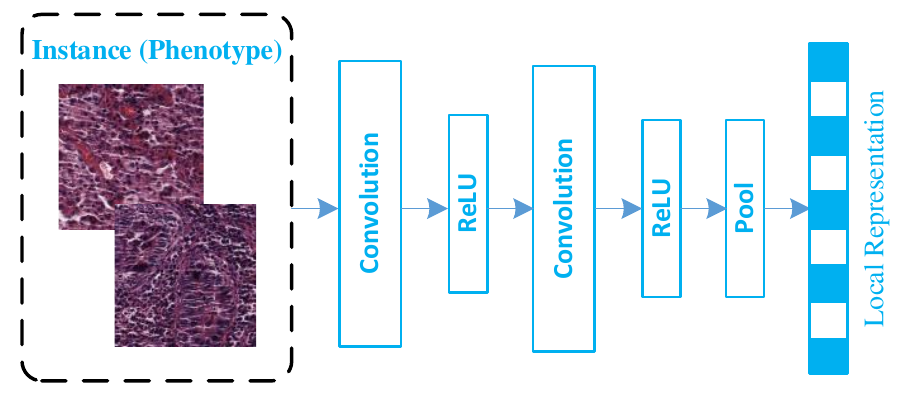}
	\caption{The network architecture in each MI-FCN. }
	\label{fig:MIL}
\end{figure}

The architecture of each Multiple Instance Fully Convolutional
Networks (MI-FCN) is shown in Fig.\ref{fig:MIL}.  
The combination of multiple layers of fully convolutional layers and non-linear activation functions has proven to be a powerful non-linear feature mapping in multiple instance problem~\cite{yang2017miml}.
The reason to use the fully convolutional networks (FCN) without including any fully connected layers is that FCN is more flexible and can handle any spatial resolution, which is needed for the considered problem since the number of patch samples in each phenotype varies.
For each phenotype, the input is a set of features from $m_i$ patches, can be organized as $1\times m_i\times d$ ($d$ is the feature dimension or channel). The network consists of several layer-pairs of $1\times 1$ conv layer and ReLU layer (we show 2 pairs in Fig.\ref{fig:MIL}). The global pooling layer (e.g. average pooling) will be added at the end. For $j$-th phenotype, its representation is denoted as $\mathbf{r_j}$.
The network receives one kind of phenotypes (tensor) as input and it can focus on local information and generate representation for the phenotype.  Since the number of patches in each phenotype varies, the fully convolutional network is more flexible to handle this scenario.


\subsubsection{Aggregation via Attention-based MIL pooling layer}
Local representations from MI-FCN encode information of the corresponding phenotype clusters and how to aggregate them into patient-level representation is one necessary step. Let $R=\{\mathbf{r}_1, \mathbf{r}_2, ..., \mathbf{r}_C\}$ be one patient with $C$ phenotype local representations and the goal is to get patient-level representation $\mathbf{z}$. The very straightforward choice is to use maximum or the mean operator, but drawbacks are very clear that they are pre-defined and non-trainable which might not be flexible and adjustable to the specific task.
Previous work~\cite{zhu2017wsisa} used weighted average of features from clusters to get the patient feature but they performed such patient-level aggregation in a separate stage and the whole approach cannot be trained end-to-end from instance-level to patient-level. A better way to integrate phenotype-level information is to leverage an attention mechanism that considers the importance of each phenotype. In this paper,
we propose to use the attention-based MIL pooling~\cite{ilse2018attention} for aggregation which is flexible and adaptive. By using such pooling operator,  the patient-level representation can be calculated as 
\begin{equation}
    \mathbf{z} = \sum_{k=1}^C a_k \mathbf{r}_k,
\end{equation} 
where
\begin{equation}
    a_k = \frac{\exp\{\mathbf{w}^\top \tanh(\mathbf{V}\mathbf{r}_k^\top) \}}{\sum_{j=1}^C \exp\{\mathbf{w}^\top \tanh(\mathbf{V}\mathbf{r}_j^\top)\}}.
\end{equation}
In the weight $a_k$ calculation, $\mathbf{w} \in \mathbb{R}^{L\times1}$ and $\mathbf{V} \in \mathbb{R}^{L \times M}$ are trainable parameters. Tangent $\tanh(.)$ element-wise non-linearity is introduced both negative and positive values for proper gradient flow. The attention-based MIL pooling allows to assign different weights to phenotype clusters within one patient and hence the final patient-level representation could be highly informative for survival prediction. In other words, it should be able to locate key clusters and provide potential ROIs.
Different from traditional attention mechanism that all instances are sequentially dependent~\cite{lin2017structured, raffel2015feed}, multiple instance learning assumes all instances are independent. As phenotype in our problem is more natural to be independent to each other, attention mechanism used in MIL pooling will be beneficial to achieve good results.

\subsubsection{Loss Function}
After attention-based MIL pooling, we will generate the patient-level aggregation from all local representations. For $i$-th patient sample passing through the proposed model, the output of this patient's hazard risk is denoted as $\mathbf{o}_i$.  Table \ref{tab:Bag_arch} presents architecture details of the proposed DeepAttnMISL. Input of our model is the set of patients' phenotype features, organized as $[(1\times m_1 \times d), (1\times m_2 \times d),..., (1\times m_c \times d)]$ where $C$ is the number of phenotypes and $m_i$ means the number of patches in $i$-th phenotype.

\begin{table}[!htb]	\caption{The architecture of DeepAttnMISL.}	
\begin{center}
		\begin{tabular}{l|l|l }
			\hline 
			{Layer} & {Input} & {Output size} \\ \hline \hline  
			MI-FCN $i$ & 1 $\times$ $m_i \times d$    & $64$ ($\mathbf{r}_i$)\\ \hline
	        Attention MIL pooling &  $64 \times C$ & 64 \\ \hline
			Fully-Con. & 64 & 32  \\ \hline
			Fully-Con. & 32 & 1 ($\mathbf{o_i}$) \\ \hline
		\end{tabular}
	\end{center}
	\label{tab:Bag_arch}
\end{table}

Denote the label of the $i$-th patient as $(t_i, \delta_i)$ where $t_i$ is the observed time, 
We assume that censoring data ($\delta = 0$, death not observed) is non-informative in that, given $\mathbf{x}_i$, the event and censoring time for the $j$-th patient are independent. Let $t_1<t_2<\dots <t_N$ denote the ordered event times. The risk set $R(t_i)$ is the set of all individuals who are still under study. For example, the patient $j$ in risk set has the survival time is equal or larger than $t_i$ ($t_j \ge t_i$). Conditioned upon the existence of a unique event at some particular time $t$ the probability that the death event occurs in the patient $i$ is
\begin{equation}
L_i = \frac{\exp(\mathbf{o}_i)}{\Sigma_{j \in R(t_i)}\exp(\mathbf{o}_j)},
\end{equation}

Assuming the patients' events were statistically independent, the joint probability of all death events conditioned upon the existence of events at those times is the partial likelihood:
\begin{equation}
L = \prod_{i:\delta_i=1}\frac{\exp(\mathbf{o}_i)}{\Sigma_{j \in R(t_i)}\exp(\mathbf{o}_j)},
\end{equation}

The corresponding log partial likelihood is
\begin{align}
l = log(L) & =  \sum_{i:\delta_i=1}(\mathbf{o}_i-\log\sum_{j:R(t_i)}\exp(\mathbf{o}_j)) = \sum_{i}\delta_i(\mathbf{o}_i-\log\sum_{j:R(t_i)}\exp(\mathbf{o}_j)),
\end{align}

The function can be maximized over the network parameters to produce maximum partial likelihood estimates. It is equivalent to minimize the negative log partial likelihood.
We then use the negative log partial likelihood as the loss function in our model as shown in below

\begin{align}\label{eq:survival_loss}
L(\mathbf{o}_i) =\sum_{i}\delta_i(-\mathbf{o}_i +   \log{\sum_{j:t_j>=t_i}\exp(\mathbf{o}_j)}).
\end{align}
 In a simplified view, the loss function contributes to overall concordance by penalizing any discordance in any values of higher risk patients if they are greater than lower those of lower risk.
Different with other deep models used the same loss function~\cite{Katzman2016deepsurv,zhu2016deep, zhu2017wsisa}, the proposed model can better fit realistic patients' whole slide imaging data and learn complex interactions using deep multiple instance representation that cover both holistic and local information.  
Since patient's risk is correlated with phenotypes from WSIs, the proposed framework can efficiently exploit phenotypes by deep multiple instance learning and attention mechanism for clinical outcome prediction at patient-level.

\section{Experiments}
\subsection{Dataset Description}

To validate the performance of the proposed DeepAttnMISL, we used two very large datasets on lung and colorectal cancers with high-resolution WSIs. 
They are the National Lung Screening Trial (NLST)~\cite{national2011national} and the Molecular and Cellular Oncology (MCO) study ~\cite{2015mco,jonnagaddala2016integration}. NLST is a very large lung cancer dataset collected by the National Cancer Institute's Division of Cancer Prevention (DCP) and Division of Cancer Treatment and Diagnosis (DCTD). The MCO study is a collection of imaging, specimen, clinical and genetic data from over 1,500 Australian individuals who underwent curative resection for colorectal cancer from 1994 to 2010. Clinical and pathological data were collected on all those cases, including follow-up data. The WSIs collection in MCO study consists of more than 1,500 WSIs representing at least one typical section from each tumour case, stained with Hematoxylin and eosin, and scanned using a 40x objective. We have different experiment comparison settings on two datasets because we only have annotations that locate tumor regions in NLST. Both datasets are good for WSI-based models as those models without requiring ROI labelling but more extensive experiments with ROI-based comparisons can only be made on NLST dataset.

The numbers of WSIs and patients in each dataset are shown in Table \ref{tab:WSIs_number}.
State-of-the-art WSI models~\cite{zhu2017wsisa, tang2019capsurv} need to control the scale of data as they will have significant computational issues on the very large number of patches. They sampled hundreds of patches per WSI and collected around 20K-200K patches in total. One advantage of the proposed model is the computational efficiency because it uses MIL with attention to aggregate 1D deep features from pre-trained models instead of training patch-based CNNs which is very time costly~\cite{hou2015efficient}.
For the purpose of training baseline WSI survival model, we first extract in total of 130K and 275K patches for MCO and NLST, respectively. We then sample more patches on MCO dataset and collect 915K patches and each WSI will have more than 500 patches. MCO study has more than 1000 patients which is much larger than data used in recent work~\cite{zhu2017wsisa, tang2019capsurv}.

\begin{table}[!htb]
    \begin{center}
        \caption{The numbers of WSIs, patients, patches, and the average number of patches per WSI extracted in each dataset.}        \label{tab:WSIs_number}
        \begin{tabular}{c|c|c|c}
            \hline 
            {Dataset} & {NLST} &{MCO\_130K} & MCO\_1M \\  \hline  
            \#patients & 387  & 1,146 & 1,146 \\ \hline
            \#WSIs & 1,177  & 1,614 & 1,614\\ \hline
            \#patches & 275,244 & 132,910 & 915,324\\ \hline
            \#patches/WSI & 234 & 82 & 567\\ \hline
        \end{tabular}
    \end{center}
\end{table}

\subsection{Implementation details}

For training, we use Adam optimization with weight decay $5 \times 10^{-4}$. The learning rate is set to $10^{-4}$ and the training monitors the loss on validation dataset and it will early stop if the loss goes increased much. 
To evaluate the performances in survival prediction, we take the concordance index (C-index) and area under curve (AUC) as our evaluation metrics~\cite{heagerty2005survival}. 
The C-index quantifies the ranking quality of rankings and is calculated as follows
\begin{equation}
c=\frac{1}{n}\sum_{i\in\{1...N|\delta_i=1\}}\sum_{t_j>t_i}I[f_i>f_j]
\end{equation}
where $n$ is the number of comparable pairs and $I[.]$ is the indicator function. $t.$ is the actual time observation. $f.$ denotes the corresponding risk. The value of C-index ranges from 0 to 1. The larger the value is, the better the model predicts.

\subsection{MCO results}
\subsubsection{Settings and Parameters}
To see effects from phenotype patterns, we tested different cluster numbers changing from 6 to 12. We split the data into 80\%  training and 20\% testing. 10\% of training data will be used as validation data for achieving early stop training. 
We would like to note that the number of phenotype clusters is the maximum number that allows each patient sample can have. The proposed model is flexible to handle patients with fewer patterns (e.g. smaller biopsy tissue). We implement this by setting the corresponding weight $a_k$ to zero if there are no patches in this cluster. To evaluate the use of different pooling ways, we built two baselines by replacing attention MIL pooling layer in DeepAttnMISL with commonly used Max and Mean pooling layer, and we indicate them as "DeepMIL+ Max/Mean" below.
Table \ref{tab:tune_phenotypes_MCO} presents results of each model. We first notice DeepAttnMISL can achieve best results in all cases which demonstrate attention MIL pooling is more flexible and better than fixed pooling operators. Second, when the phenotype is set to large values, results get worse which show more clusters actually cannot guarantee prediction benefits. 
\begin{table}[!htb]	\caption{ Performances with different number of phenotypes.}
	\begin{center}
		\begin{tabular}{ c|cccc}\hline
			{\bfseries Model } & c=6 & c=8 & c=10 & c=12  \\ \hline
			DeepAttnMISL & \textbf{0.652} & \textbf{0.648} & \textbf{0.624} & \textbf{0.607}  \\ 
			DeepMIL + Max & 0.594 & 0.606 & 0.606 & 0.540\\
			DeepMIL + Mean & 0.604 & 0.578 & 0.604 & 0.601 \\ \hline
		\end{tabular}
	\end{center}
	\label{tab:tune_phenotypes_MCO}
\end{table}

The basic MI-FCN network of our DeepAttnMISL consists of one convlolutional layer, one ReLU layer, one pooling layer. We study the effects of different number of convolution and ReLu layer-pairs and report results in Table \ref{tab:tune_layers_MCO_cur}. For the 1 layer, we used 64 filters in the convlolutional layer. We used $\{2048, 64\}$ number of filters in 2 layers and $\{2048, 1024, 64\}$ for 3 layers setting, respectively. From the table, we decide to choose one convolutional-ReLU layer pair with Global Average Pooling in MI-FCN network.
\begin{table}[!htb]\caption{ Results under different network configurations on testing data. The cluster number is set to 6.}
	\begin{center}		
		\begin{tabular}{ c|ccc}\hline
			{c=6 } & 1 layer & 2 layers & 3 layers  \\ \hline
			Global Average Pooling & \textbf{0.652} & 0.634 & 0.644   \\ 
			Global Max Pooling & 0.615 & 0.640 & 0.623 \\ \hline
		\end{tabular}
	\end{center}
	\label{tab:tune_layers_MCO_cur}
\end{table}

To validate the effectiveness of Siamese, we then remove the Siamese network and only use attention pooling layer on input features. In this case, no phenotype clusters are considered. This scenario will be the direct application of attention aggregation without using phenotype clusters~\cite{ilse2018attention}. 5 fold cross-validation is performed with the cluster number 6 on MCO-130K. Results can be found in Table \ref{tab:siamese}. We can see the overall performance is not good as the DeepAttnMISL which means the importance of Siamese network. This validates the effectiveness of phenotype clusters in Siamese network.
The final c-index across 5 folds is $0.542\pm 0.022$ for model without Siamese and $0.595 \pm 0.036$ for model with Siamese, respectively. Results suggest the usefulness of phenotype patterns and the Siamese architecture.

\begin{table}[!htb]\caption{ Validation of Siamese on MCO-130K dataset.}
	\begin{center}		
		\begin{tabular}{ c|ccccc}\hline
			{c=6} & fold 1 & fold 2 & fold 3 & fold 4 & fold 5    \\ \hline
			No Siamese & 0.564 & 0.538 & 0.515 & 0.527 & 0.564    \\
			w Siamese & 0.652 & 0.579 & 0.609 & 0.564 & 0.573  \\ \hline
 			
		\end{tabular}
	\end{center}
	\label{tab:siamese}
\end{table}

To validate effects of different components, we add more evaluations by changing encoder/clustering part, and results can be found in the Table \ref{tab:InceptionV3}. The more advanced InceptionV3~\cite{szegedy2016rethinking} model is tested and we also introduce spectral clustering as the alternative method for Kmeans. All other settings and architectures are kept the same. Details about each fold can be seen in the Table \ref{tab:InceptionV3}. For model with InceptionV3 and Kmeans clustering, C-index result is $0.598 \pm 0.054 $ on 5-fold cross validation. When changing Kmeans clustering to spectral clustering, the performance is $0.593 \pm 0.032$. Compared with the model using VGG-16 and Kmeans clustering ($0.595 \pm 0.036$), performances from different variants of models are quite similar. Therefore, we decide to use VGG-16 and Kmeans clustering for comparisons.

\begin{table}[!htb]\caption{ Results with different feature extractor and clustering on MCO-130K dataset.}
	\begin{center}		
		\begin{tabular}{ c|ccccc}\hline
			{c=6 } & fold 1 & fold 2 & fold 3 & fold 4 & fold 5   \\ \hline
			InceptionV3+k & 0.670 & 0.540 & 0.630 & 0.603 & 0.549  \\ 
			InceptionV3+sp & 0.630 & 0.565 & 0.611 & 0.554 & 0.607  \\ \hline
		\end{tabular}
	\end{center}
	\label{tab:InceptionV3}
\end{table}

We also try with the more advanced gating mechanism~\cite{ilse2018attention,dauphin2017language} together with $tanh(.)$ non-linearity in eq (2). Results on MCO-130K are reported in Table \ref{tab:gated}. We can find gated-attention and plain attention mechanism behave similarly in different phenotype cluster settings but the plain attention is slightly better.
 

\begin{table}[!htb]\caption{ Results of different attention mechanisms on MCO-130K dataset.}
\begin{center}	
\begin{tabular}{l|cc}
\hline
     & Gated-Attention     & Attention           \\ \hline
c=6  & 0.596 (0.029) & 0.595 (0.036) \\ \hline
c=8  & 0.586 (0.043) & 0.599 (0.049) \\ \hline
c=10 & 0.561 (0.048) & 0.585 (0.036) \\ \hline
c=12 & 0.579 (0.031) & 0.591 (0.026) \\ \hline
\end{tabular}
\end{center}
\label{tab:gated}
\end{table}

\subsubsection{Comparisons}

WSISA~\cite{zhu2017wsisa} is one representative WSI-based survival learning but it only extracts features from WSIs and needs a separate survival learning to get final predictions. We choose three top survival models according to settings in WSISA~\cite{zhu2017wsisa}, they are Lasso-Cox~\cite{tibshirani1997lasso}, En-Cox~\cite{yang2012cocktail} and MTLSA~\cite{li2016multi}. As WSISA has the computational issue when there are too many patches in the whole dataset and thus the scale of 100K-200K patches is acceptable for experiments. 
We have a another collection of patches with around 1 million patches to see effects from the patch scale but only perform our model on this scale because training with WSISA is not endurable.

\textcolor{red}{Our preliminary work DeepMISL~\cite{yao2019deep} has shown the effectiveness of using both global and local representation from Multiple Instance Learning can benefit survival prediction. However, the model still treats phenotype clusters equally and cannot recognize clusters that contribute more on patients' survival.}
We perform 5 fold cross-validation and report the average values of C-index and AUC on Table \ref{tab:MCO_CI} and \ref{tab:MCO_AUC}, respectively. From both tables, one can see that the proposed method achieves best results than models using WSISA features in all cluster number settings on MCO-130k. Improvements can be related to the following differences.  
First, clustering is performed on patient-wise while recent WSI-based approaches~\cite{zhu2017wsisa, tang2019capsurv} need to cluster on all patches from patients of the database. 
Because WSISA~\cite{zhu2017wsisa} needs independent DeepConvSurv to select important clusters and it has to divide the whole dataset into different types by clustering on all patches.
\textcolor{red}{DeepMISL~\cite{yao2019deep} can combine both local and bag representation with MIL but it is still unable to treat phenotype clusters differently which will limit its use on larger datasets.}
With the advantage of MIL and attention mechanism, the proposed DeepAttnMISL can easily find important instances (clusters) within the bag are more likely to achieve better patient-level predictions. There is no need to perform clustering on the whole dataset. A trainable and adaptive attention-based MIL pooling in DeepAttnMISL can adjust to a task and data which could help succeed in calculating the better patient representation.  Increases with 1\%-3\% are observed when we use more patches from MCO-1M data and this reminds us more patches can benefit predictions but actually cannot offer significant improvements. This demonstrates the robustness of the proposed DeepAttnMISL that is not rely on the number of sampling patches.

\begin{table}[!htb]\caption{ C-index values of the proposed model and WSISA with different settings.}
\begin{center}	
\begin{tabular}{l|ccccc}
\hline
Method                        & Settings & c=6 & c=8 & c=10 & c=12 \\ \hline
\multirow{2}{*}{DeepAttnMISL} & 130K     &  0.595   & 0.599    & 0.585  &  0.591     \\
                              & 1M       & \textbf{0.606}    & \textbf{0.600}    &   \textbf{0.603}   &  \textbf{0.599}    \\ \hline

\multirow{2}{*}{DeepMISL} & 130K & 0.557 & 0.547 & 0.587 & 0.543 \\
& 1M & 0.569 & 0.575 & 0.573 & 0.567 \\ \hline
W-MTLSA                   & 130K     &  0.558   &  0.567   &  0.524    &  0.547   \\
W-LassoCox                & 130K     & 0.552    &  0.546   &  0.503    &  0.523    \\
W-EnCox                   & 130K     &  0.552   &  0.545   &  0.504    &  0.522 \\ \hline   
\end{tabular}
\end{center}
\label{tab:MCO_CI}
\end{table}

\begin{table}[!htb]\caption{ AUC values of the proposed model and WSISA with different settings.}
\begin{center}	
\begin{tabular}{l|lllll}
\hline
Method                        & Settings & c=6 & c=8 & c=10 & c=12 \\ \hline
\multirow{2}{*}{DeepAttnMISL} & 130K     &  0.623   & \textbf{0.640}    & \textbf{0.636}  &  0.622     \\
                              & 1M       &  \textbf{0.644}   & 0.638    &   0.633   &  \textbf{0.637}    \\ \hline
\multirow{2}{*}{DeepMISL} & 130K & 0.564 & 0.552 & 0.590 & 0.547 \\
& 1M & 0.570 & 0.587 & 0.579 & 0.576 \\ \hline
W-MTLSA                   & 130K     &  0.560   &  0.560   &  0.531    &  0.555    \\
W-LassoCox                & 130K     &  0.531  &  0.541   &  0.495    &   0.495   \\
W-EnCox                   & 130K     &  0.532   &  0.544   &  0.497    & 0.496 \\ \hline   
\end{tabular}
\end{center}
\label{tab:MCO_AUC}
\end{table}

\begin{figure}[!htb]
	\centering
	\includegraphics[width=0.8\linewidth]{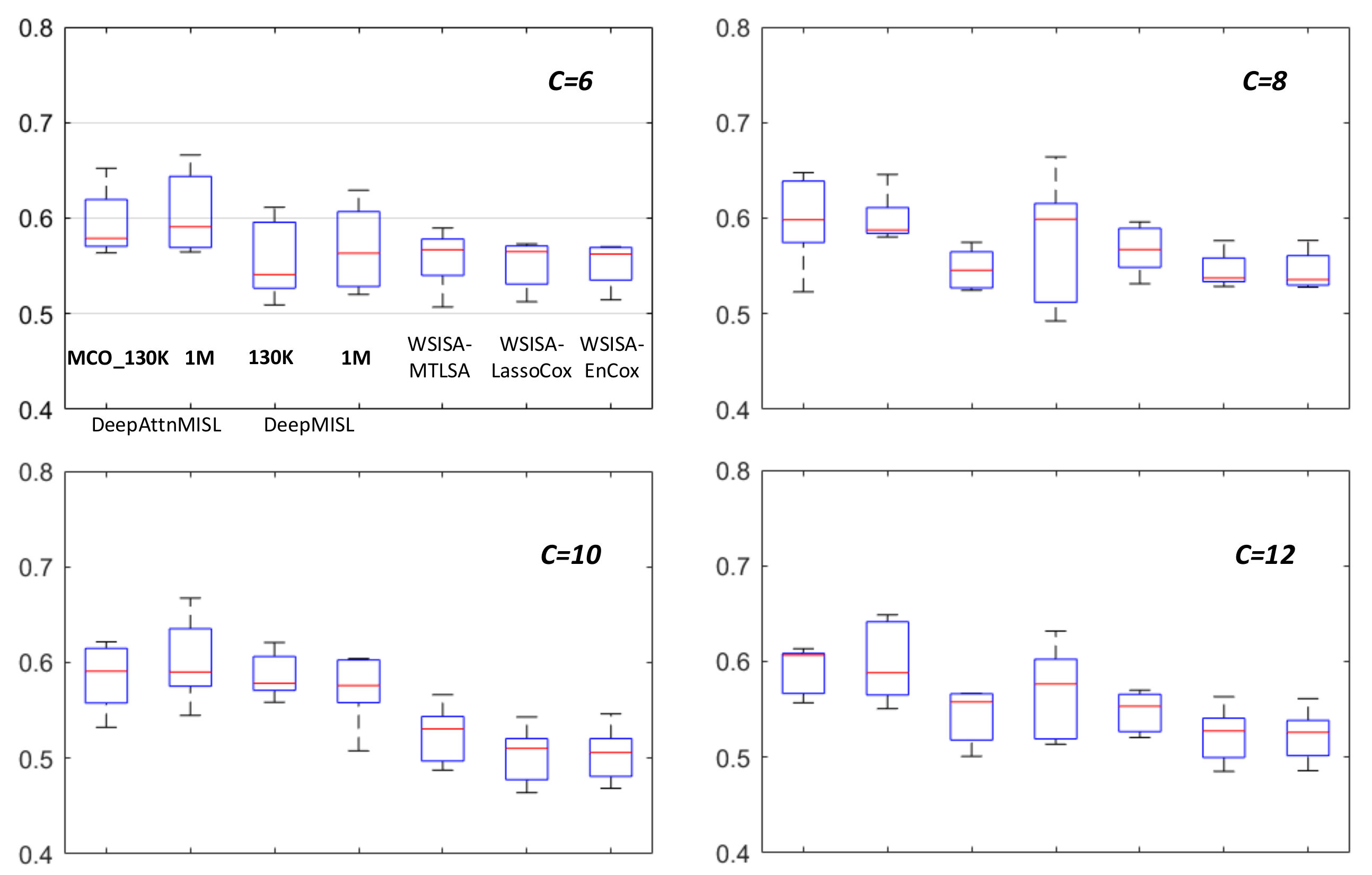}
	\caption{Boxplots of C-index values with different numbers of phenotype patterns.}
	\label{fig:MCO_ci_boxplot}
\end{figure}

\begin{figure}[!htb]
	\centering
	\includegraphics[width=0.8\linewidth]{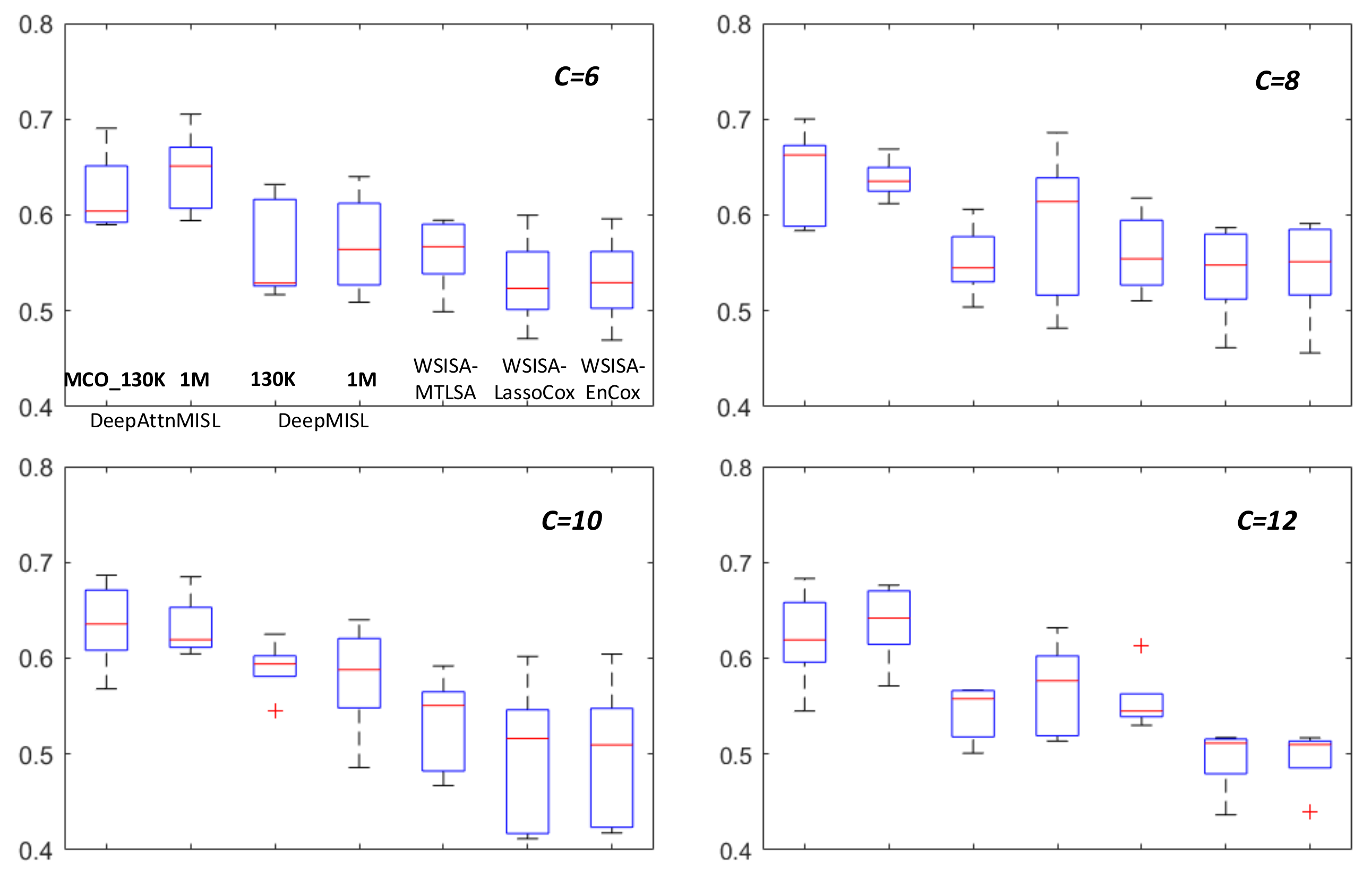}
	\caption{Boxplots of AUC values with different numbers of phenotype patterns.}
	\label{fig:MCO_auc_boxplot}
\end{figure}

Fig.\ref{fig:MCO_ci_boxplot} and Fig.\ref{fig:MCO_auc_boxplot} present boxplots of C-index and AUC values from each model with different phenotype cluster numbers. We only show captions in the top left figure and others will also share this description. We can see that results of our method on MCO-1M and MCO-130K don't have significant differences. This shows sampling strategies will not affect final results of the proposed method in cross-validation settings. 
One can observe that our models consistently perform better than WSISA models across different phenotype cluster numbers.

\begin{figure}[!htb]
	\centering
	\includegraphics[width=0.8\linewidth]{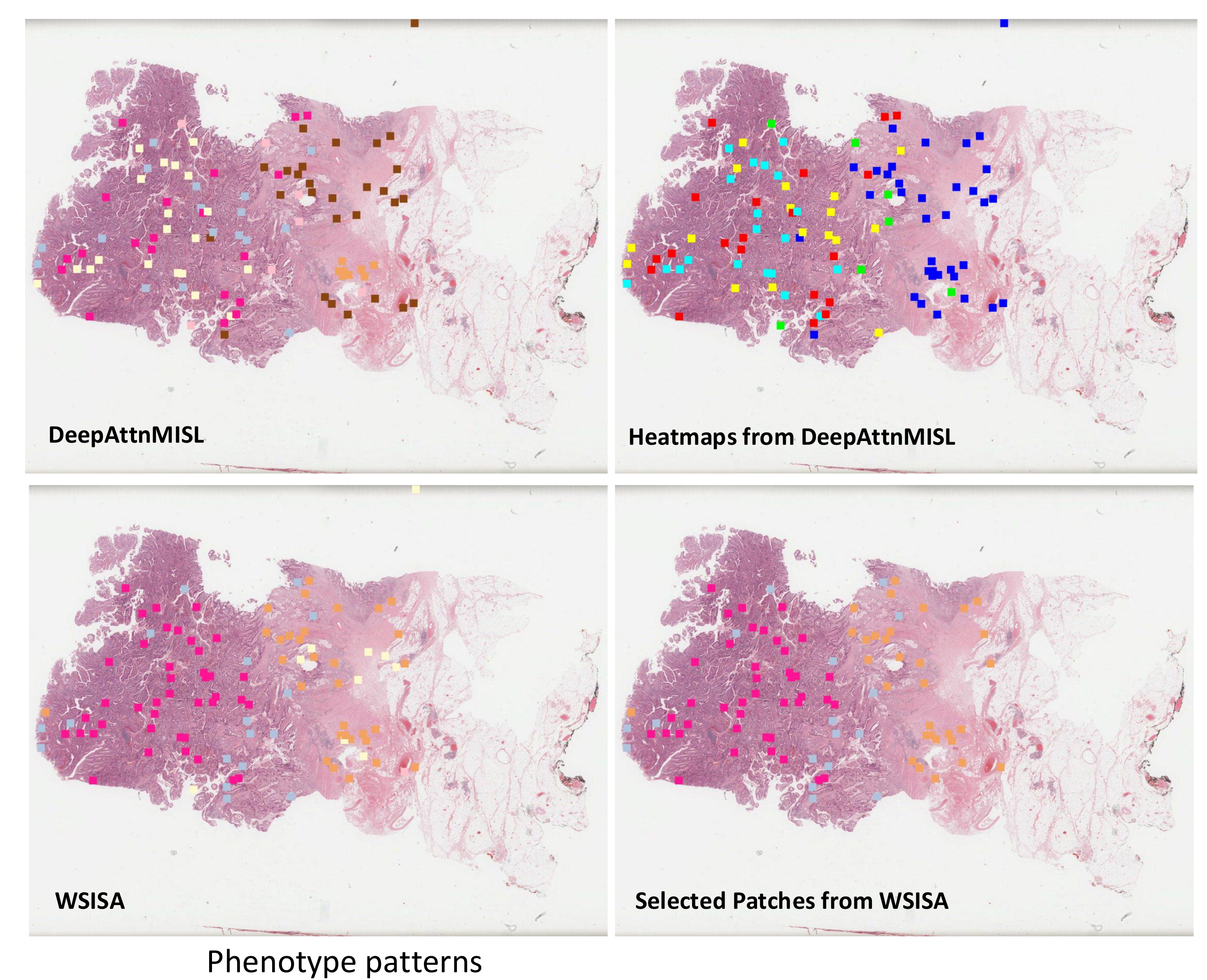}
	\caption{Comparison of phenotype patterns distribution in the first column. The second column shows heatmap and selected patches from the proposed model and WSISA on MCO-130K, respectively.}
	\label{fig:MCO_tune_c}
\end{figure}

Fig.\ref{fig:MCO_tune_c} visualizes clustered phenotype patterns and selected patches from DeepAttnMISL and WSISA on MCO-130K when cluster number is set as 6. The first row shows results from DeepAttnMISL while the second one presents results from WSISA.
In MCO-130K, around 100 patches per WSI are sampled and it clearly can see that clustering based on VGG-16 features is capable of identifying patches from different layers of WSI and grouping similar patches into the same category. The most important advantage of DeepAttnMISL is its good interpretability and we create a heatmap by showing the corresponding attention weight of each phenotype cluster. We rescaled the attention weights using $a_k^{'} = (a_k - min(a))/(max(a)-min(a))$.
Red color indicates the highest attention weight while blue means the lowest values. From the obtained heatmap, we can see the proposed approach can identify higher risk regions properly because most of patches with high attention weights are from tumor regions. When we look at selected patches from WSISA, we can observe that many patches from non-tumor regions are also selected. That is because WSISA selects clusters based on patches from the whole database and thus it cannot guarantee reliable selection on the specific patient due to the heterogeneity across patients.

\begin{figure}[!htb]
	\centering
	\includegraphics[width=0.8\linewidth]{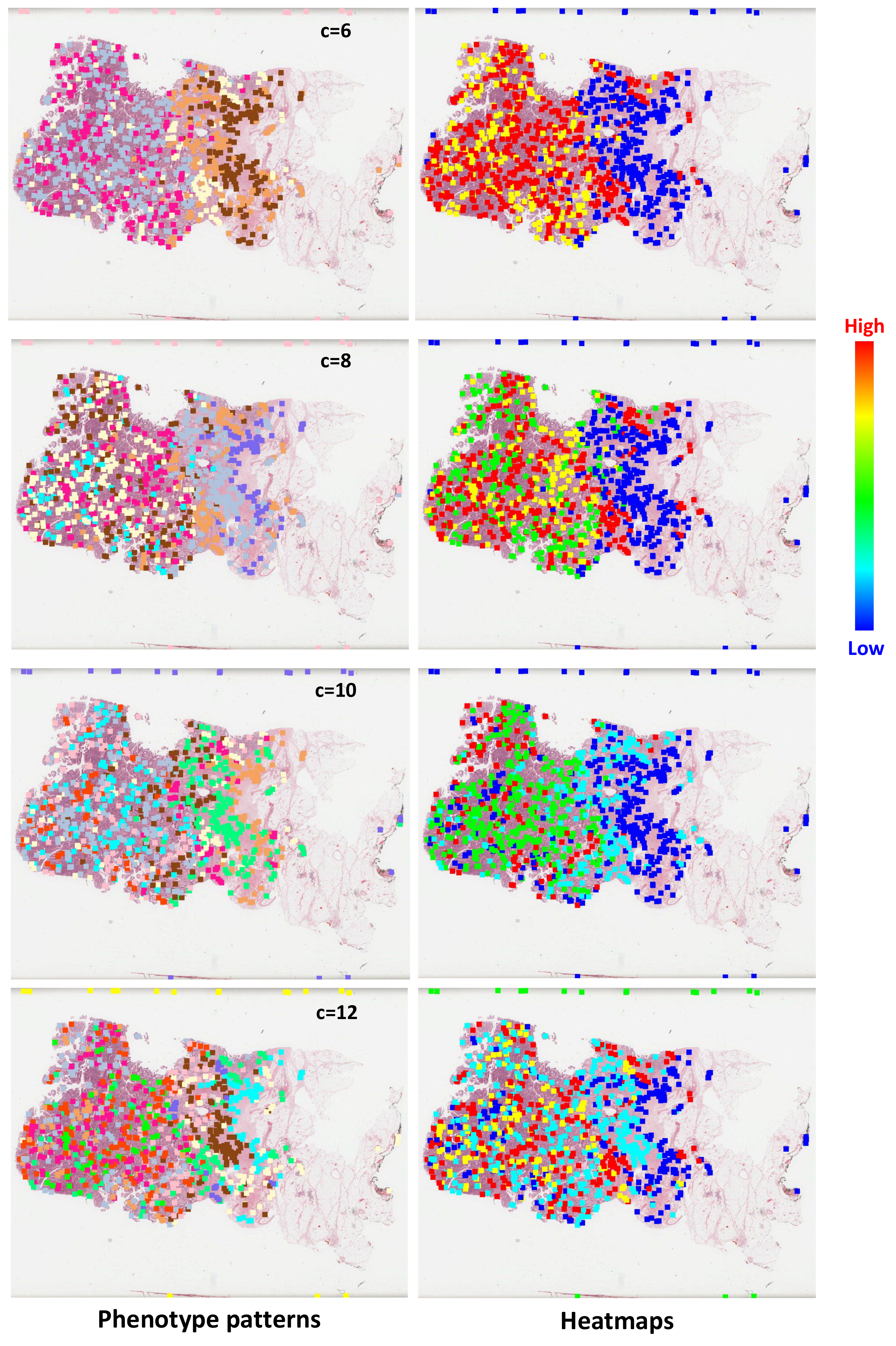}
	\caption{Phenotype patterns clustering visualizations and the corresponding heatmaps from the proposed model on MCO-1M.}
	\label{fig:MCO_1M}
\end{figure}

More clear visualizations can be found in Fig.\ref{fig:MCO_1M} on MCO-1M set and more patches (about 1000) are sampled per WSI. The first column shows phenotype patterns from the proposed model with different numbers. The second column shows the corresponding heatmaps. Attention mechanism in DeepAttnMISL allows to easily interpret the provided decision in terms of instance-level labels. From heatmaps, we can see results from $c=6$ and $c=8$ look better as most patches from cancerous regions are given by high attention weights.

\begin{figure}[!htb]
	\centering
	\includegraphics[width=0.8\linewidth]{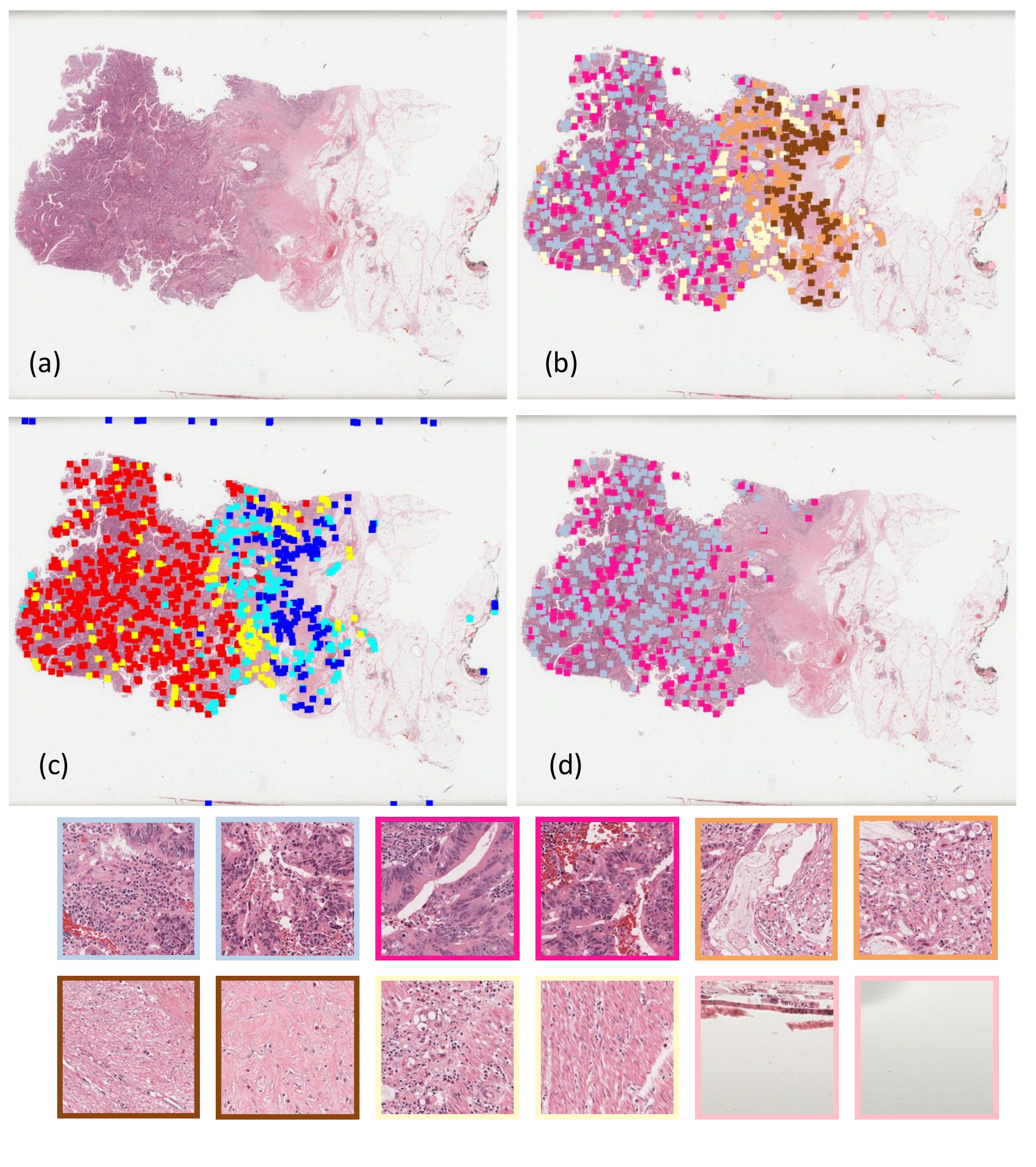}
	\caption{(a) Original WSI, (b) Phenotype patterns distribution, (c) Heatmaps from our model, (d) Selected patches with highest attention weights. The bottom shows representative patches from each phenotype.}
	\label{fig:MCO_vis1}
\end{figure}
\begin{figure}[!htb]
	\centering
	\includegraphics[width=0.7\linewidth]{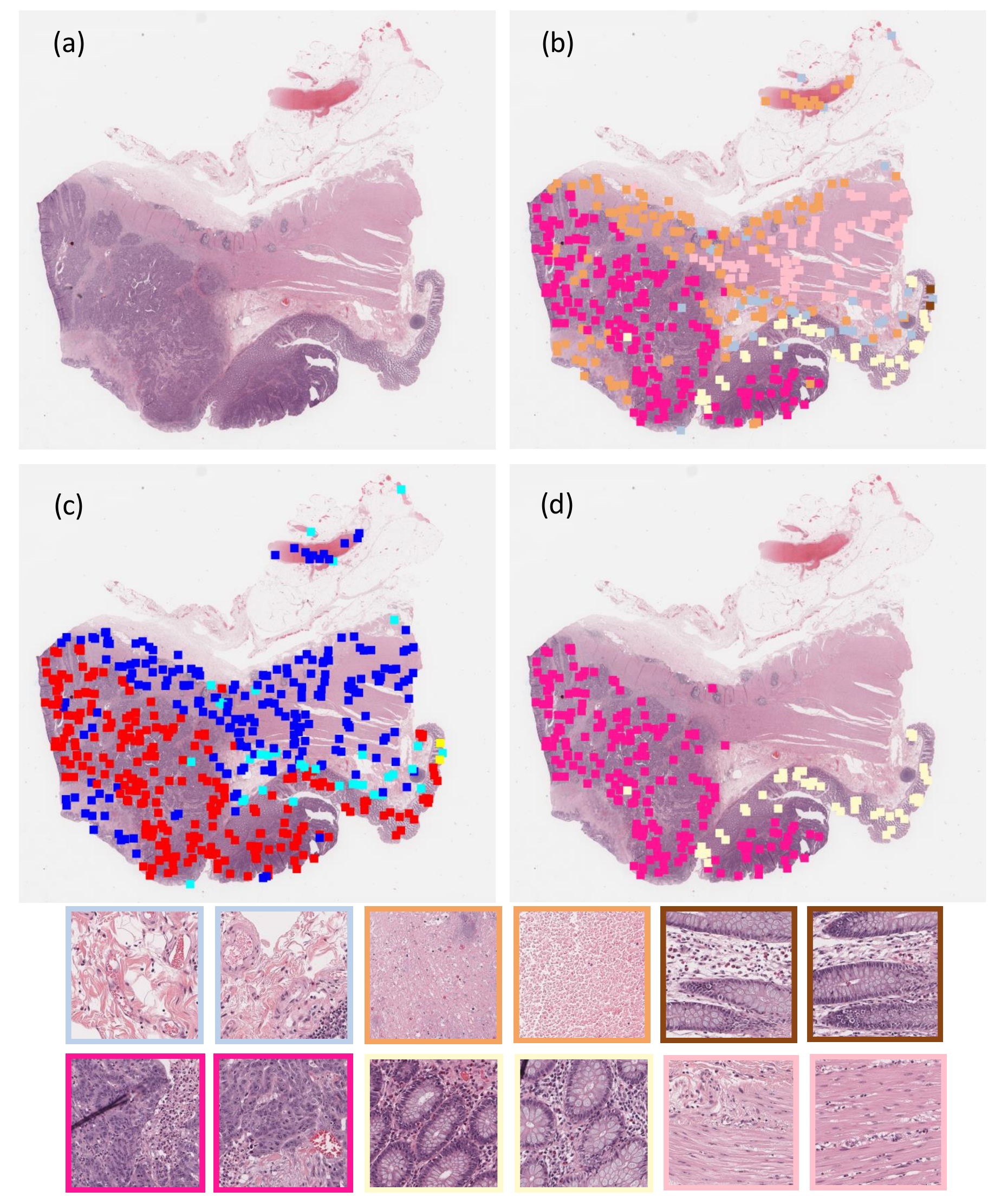}
	\caption{(a) Original WSI, (b) Phenotype patterns distribution, (c) Heatmaps from our model, (d) Selected patches with highest attention weights. The bottom shows representative patches from each phenotype.}
	\label{fig:MCO_vis2}
\end{figure}

To better visually validate the effect of attention mechanism, we collect and examine the attention weights as well as their corresponding patch images on MCO-1M data in Fig.\ref{fig:MCO_vis1}.  The bottom shows randomly selected patches from each phenotype and the frame colors of patches correspond to pattern colors in Fig.\ref{fig:MCO_vis1}-(b). We use threshold as 0.8 to only show patterns with higher attention weights in Fig.\ref{fig:MCO_vis1}-d. Each color represents each phenotype pattern of the whole slide image and we can see the proposed model has higher interest on patches more related to tumor regions. Relative low attention weights are given to normal tissue regions. More surprisingly, the model can also give low attentions on background regions as they don't provide any information and are noisy images.
Fig.\ref{fig:MCO_vis2} shows another example. From the figure, we can see most patches from tumor regions are found and our model can successfully assign higher attention weight for such pattern. For patches with relatively less complex structures and textures, our model can identify them as not very important regions by giving lower attention weights.   

\subsection{Lung Cancer dataset results}
\subsubsection{Baseline models}
As we have annotations on NLST dataset, we can conduct more extensive experiments with ROI-based survival models.
Following the recent framework~\cite{yu2016predicting}, we extracted 10 dense image patches from ROIs and calculated hand-crafted features using CellProfiler~\cite{carpenter2006cellprofiler} which serves as a state-of-the-art medical image feature extracting and quantitative analysis tool.  A total of 1,795 quantitative features were obtained from each image tile. Then we averaged those features across different patches for each patient.
These types of image features include cell shape, size, texture of the cells and nuclei, as well as the distribution of pixel intensity in the cells and nuclei. 
We can summarize the comparison methods into five categories as follows:
\begin{itemize}
	\item \textbf{Cox models}: The Cox proportional hazards model is the most commonly used semi-parametric model in survival analysis. Two regularized Cox models $l_1$-norm (LASSO-Cox)~\cite{tibshirani1997lasso} and boosting cox model (Cox-boost)~\cite{Binder2008Allowing} are compared in experiments.
	\item \textbf{Parametric censored regression models}: PCR models formulates the joint probability of the uncensored and censored instances as a product of death density function and survival functions, respectively~\cite{lee2003statistical}. We choose Weibull, Logistic distribution to approximate the survival data. 
	\item \textbf{MTLSA}: Multi-Task Learning model for Survival Analysis (MTLSA)~\cite{li2016multi} reformulates the survival model into a multi-task learning problem.
	\item \textbf{WSISA}: WSISA can learn effective features from WSIs~\cite{zhu2017wsisa}. We train LassoCox and MTLSA using WSISA learned features as they are top models based on their report. To investigate performance from pre-trained network, a ResNet34~\cite{he2016deep} model is used and then fine-tuned as the backbone network in WSISA.
	
	\item \textbf{DeepMISL}: Deep Multiple Survival Learning combined both local and global representation to predict outcomes~\cite{yao2019deep}. 

\end{itemize}

\subsubsection{Results}

We reported results from a few possible numbers of phenotypes, such as $\{6, 8, 10, 12\}$ on the testing dataset. From the Table \ref{tab:tune_phenotypes}, we can see models using fewer clusters are unable to achieve good results. The reason might be patches of lung cancer patients are very heterogeneous and it is relative difficult to learn survival-related representations from fewer phenotypes.
Results suggest the number of 10 achieves slightly better predictions which is consistent with findings in WSISA~\cite{zhu2017wsisa}. Thus, we decide to choose to cluster 10 phenotypes in our model. Other parameters are kept the same with settings in MCO experiments.

\begin{table}[!htb]	\caption{ Performances with different number of phenotypes.}
	\begin{center}
		\begin{tabular}{ c|cccc}\hline
			{\bfseries No.} & 6 & 8 & 10 & 12 \\ \hline
			CI & 0.673 & 0.769 & 0.775 & 0.742  \\ \hline
		\end{tabular}
	\end{center}
	\label{tab:tune_phenotypes}
\end{table}



Table~\ref{tab:ci_NLST} shows C-index and AUC values by various survival regression methods on 5-fold cross validation. It shows the prediction power of the proposed method compared with different survival models. One can see that the proposed method achieves both highest C-index and AUC values which present the best prediction performance among all methods.
From the table, baseline models using hand-crafted features perform not well due to following reasons: 1) the limitation of local information provided by the patches extracted from the ROI using hand-crafted features; 2) the non-effective aggregation way to represent the heterogeneity of tumor and patient from patch-based results. Instead of using a small set of patches and human-designed features, the proposed method can effectively learn complex deep bag representation from phenotype patterns to predict patient survival outcomes.

\begin{table*}[!htb]	\caption{Performance comparison of the proposed methods and other existing related methods using C-index values on NLST dataset.}
\centering
\begin{tabular}{p{3cm}|p{7cm}|p{2.5cm}|p{2.5cm}}
\hline
Type    & Method       & C-index & AUC   \\ \hline
\multirow{4}{*}{Deep Learning} & DeepAttnMISL &    \textbf{0.6963} (0.0660)      &  \textbf{0.7143} (0.0541)    \\
& DeepMISL~\cite{yao2019deep} & 0.6476 (0.0698)      &  0.6693 (0.0866)  \\
& Finetuned-WSISA-LassoCox \cite{zhu2017wsisa}        &   0.6123 (0.0216)    &   0.6427 (0.0575)  \\
                                                                          & Finetuned-WSISA-MTLSA \cite{zhu2017wsisa} & 0.6428 (0.0259) & 0.6963 (0.0668) \\

                                                                          & WSISA-LassoCox \cite{zhu2017wsisa}        &   0.5996 (0.0750)    &   0.5957 (0.0674)  \\
                                                                          & WSISA-MTLSA \cite{zhu2017wsisa} & 0.6305 (0.0575) & 0.6479 (0.0936) \\ \hline
\multirow{2}{*}{Cox-based}                                                & Lasso-Cox~\cite{tibshirani1997lasso}    &    0.4842 (0.0508)     &  0.4903 (0.1011)   \\
        & Cox-boost~\cite{Binder2008Allowing} & 0.5474 (0.0370) & 0.5271 (0.0386) \\ \hline
\multirow{2}{*}{Parametric models}                                        & Logistic~\cite{kalbfleisch2011statistical}     &    0.4998 (0.0881)     &  0.5013 (0.1146)   \\
                                                                          & Weibull~\cite{kalbfleisch2011statistical}       &    0.5577 (0.0395)     &   0.5618 (0.0976)  \\ \hline
Multi-task based                                                          & MTLSA~\cite{li2016multi}        &  0.5053 (0.0509) &  0.5362 (0.0416)    \\ \hline
Ranking based                                                             & BoostCI~\cite{mayr2014boosting}      &    0.5595 (0.0610)     &  0.5487 (0.0532) \\ \hline
\end{tabular}

	\label{tab:ci_NLST}
\end{table*}

WSISA achieves better results than baseline models which shows the good representative ability of features from WSISA.
However, WSISA needs a separate stage to train several DeepConvSurv models independently and will discard some phenotypes in the final stage, the performance actually depends on how well to select important clusters and WSISA still has the chance to lose in selecting survival-related clusters for a good final survival prediction.
To investigate results from pre-trained model, we replaced the original 2DCNN network of WSISA by using a pre-trained ResNet34~\cite{he2016deep}. The whole model will be fine-tuned following the same process. It is clear to see improved C-index which can demonstrate that fine-tuned models can bring benefits but our DeepAttnMISL is still better than finetuned-WSISA by a large margin.
When introducing MIL into survival learning, DeepMISL and the proposed model can improve predictions from WSISA by C-index metric. Instead of selecting phenotypes, DeepMISL and the proposed model are designed to consider all possible patterns. Performance is further improved when we use the more flexible attention mechanism to learn informative and discriminate patterns. This architecture makes the proposed method can better learn heterogeneous information encoded in WSIs which will make it more practical and have better intepretability than DeepMISL in real applications.

\begin{figure}[!htb]
	\centering
	\includegraphics[width=0.85\linewidth]{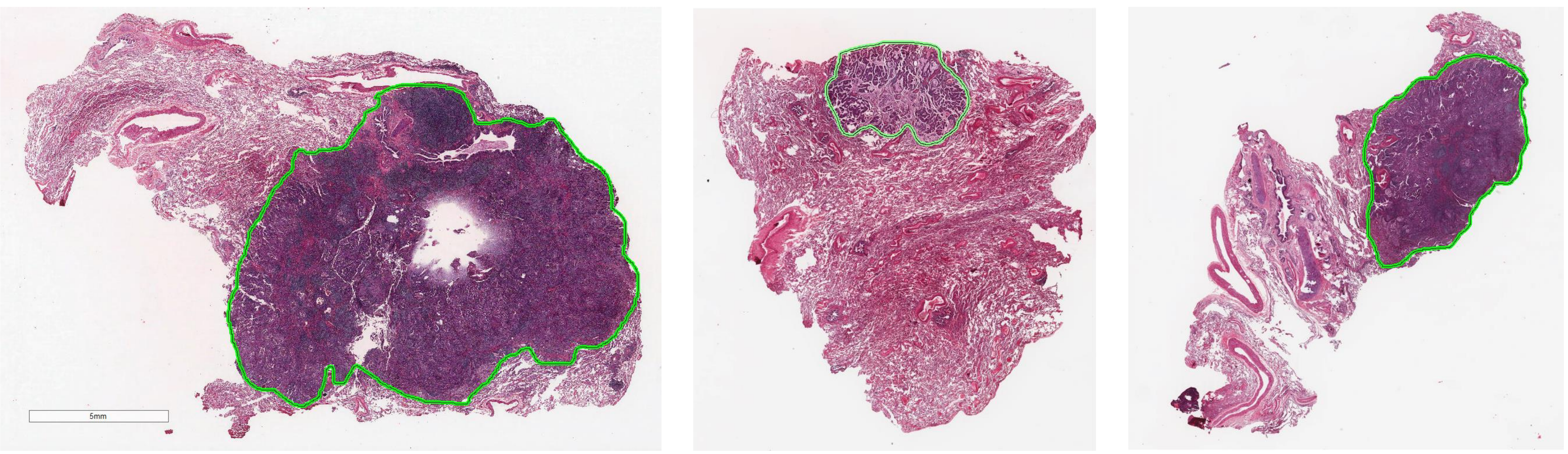}
	\caption{WSI Annotations of one example patient.}
	\label{fig:nlst_rois}
\end{figure}

\begin{figure*}[!htb]
	\centering
	\includegraphics[width=0.85\linewidth]{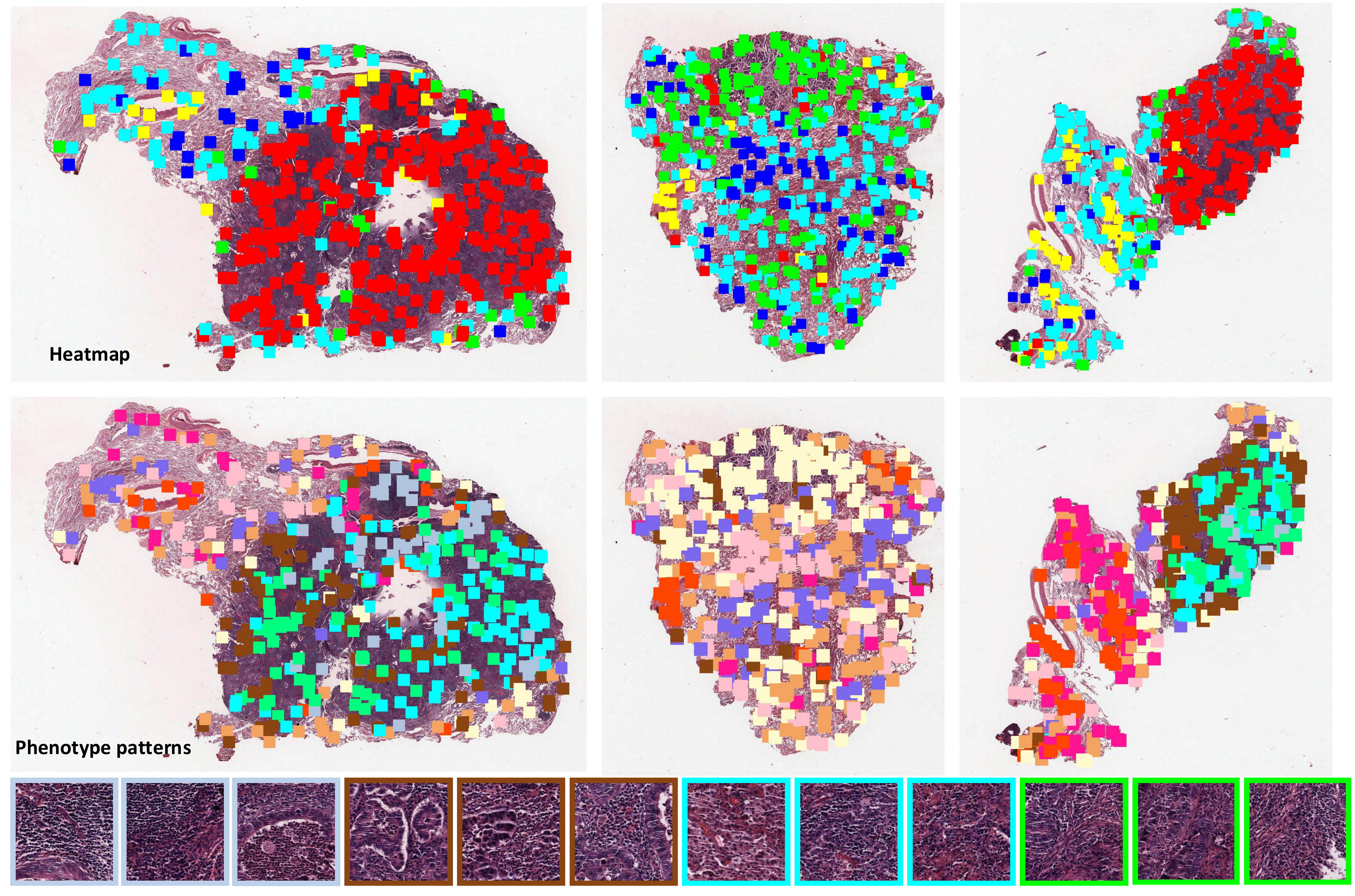}
	\caption{Phenotype pattern distribution and the corresponding heatmaps from the proposed model on three WSIs of the same patient. The bottom shows patches from phenotypes with high attention values.}
	\label{fig:nlst_vis2_ours}
\end{figure*}

\begin{figure*}[!htb]
	\centering
	\includegraphics[width=0.8\linewidth]{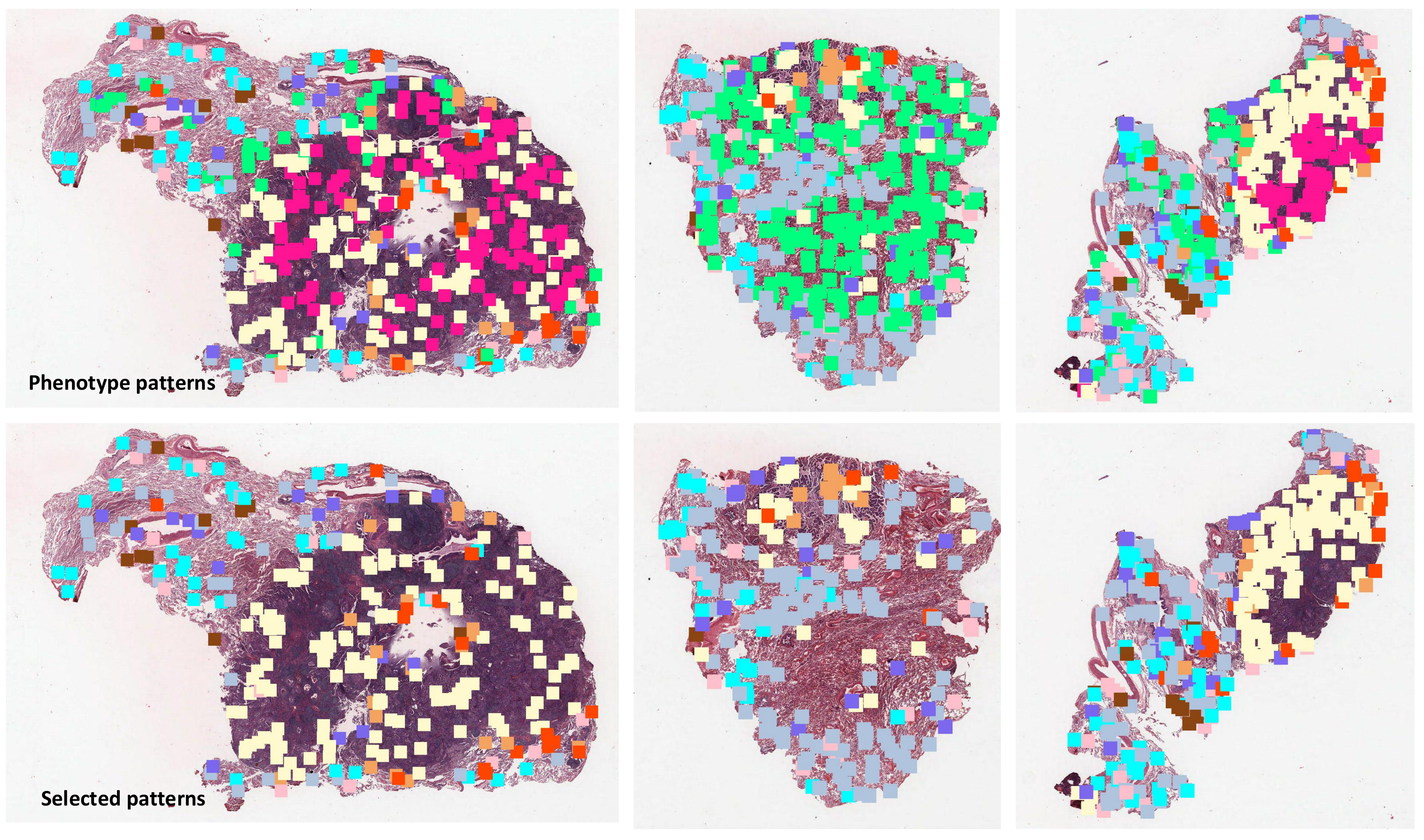}
	\caption{Phenotype pattern distribution and selected patterns from WSISA. Missing tumor patches can be observed from selected patterns by WSISA.}
	\label{fig:nlst_vis2_WSISA}
\end{figure*}

We pick one patient as the example to show visualization results. Fig.\ref{fig:nlst_rois} presents this patient's all WSIs and the corresponding tumor region annotations.
Fig.\ref{fig:nlst_vis2_ours}-\ref{fig:nlst_vis2_WSISA} show results from the proposed model and WSISA, respectively. In Fig.\ref{fig:nlst_vis2_ours}, the first row shows attention weights heatmaps and the second row shows phenotype pattern distributions on original WSIs. The bottom presents randomly selected patches with higher attention weights (patches with red colors in heatmaps). It is clear to see that most patches from tumor regions are highlighted with high attentions while patches from normal tissues are treated with lower attentions. Compared with results from WSISA shown in Fig.\ref{fig:nlst_vis2_WSISA}, we can see that WSISA will miss many tumor patches and select many normal patches as discriminative patterns. Patches from cancerous regions can be grouped in similar clusters but not all of them will be selected by WSISA as the selection is performed via DeepConvSurv on all patches of the database. Selected phenotypes are more likely discriminative for the whole database with all patients and they are not well interpreted for the specific patient.

\begin{figure*}[!htb]
	\centering
	\includegraphics[width=1\linewidth]{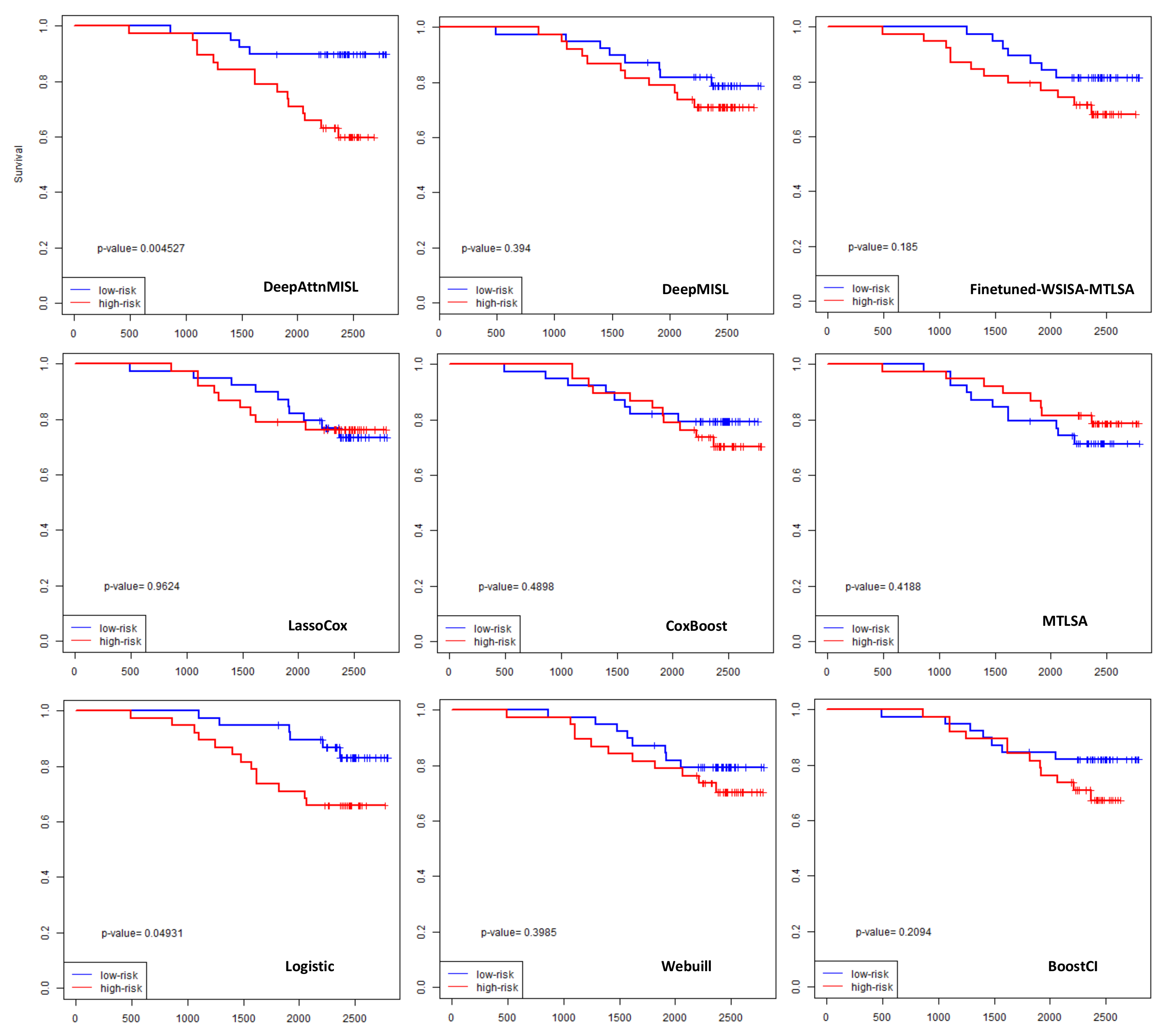}
	\caption{Kaplan-Meier survival curves of different models for one testing fold. High risk (great than median) groups are plotted as green lines, and low risk (less than or equal to median) groups are plotted as red lines. The x axis shows the time in days and y axis presents the probability of overall survival. Log rank $p$ value is shown on each figure.  "+" means the censored patient.}
	\label{fig: kmplot}
\end{figure*}

Given the trained survival models, we can use the estimated testing risk scores to classify patients into low or high-risk group for personalized treatments. Two groups are classified by the median of predicted risk scores. We evaluate if those models can correctly classify death patients (uncensored data) into two groups since uncensored data is more informative. Patients with longer survival time should be classified into low risk group and vice versa. If the model cannot correctly distinguish high and low risk death patients, two average death times should be very close. We plot Kaplan-Meier survival curves on one testing fold in Fig.\ref{fig: kmplot}. From the figure, one can see that the proposed model can more successfully group testing death patients into two groups than other methods in all datasets.
The log rank test is conducted to test the difference of two curves. It is shown that the proposed method can achieve the most significant log rank test outcome (p-value = $4.527 \times 10^{-3}$) while some of others do not reach statistical significances. Kaplan-Meier curves suggest that the proposed comprehensive prediction model can offer personalized risk scores which can better group individuals into two groups. The proposed model has a significant impact on population survival times. It can be used as a recommendation system for offering personalized treatments by determining the relationship between a patient's whole slide pathological images and his or her risk of an event (death). 

p value of log-rank test less than 0.05 is considered significant and $p<0.1$ is marginal significant. For five testing folds, our model can achieve four significant results and one marginal significant result with $p=0.09$, DeepMISL achieves two significant results and one marginal significant with $p=0.07$, Finetuned-WSISA-MTLSA achieves one significant result and three marginal significant results. Logistic model achieves only one significant result ($p=0.049$), shown in Fig.13.
BoostCI has one fold significant result ($p=0.0123$). All other baseline models cannot have significant results on all testing folds.
Overall, deep learning
models perform well to achieve significant results than models with hand-crafted features. The proposed model achieves the most number of significant results which could validate that the predictor from our model is an important prognostic factor which could be used for a good patient risk stratification.

\subsection{Ensemble Models}
We investigated if ensemble models could benefit final results. During each fold, we train five models and then average prediction score on the corresponding testing fold. The maximum cluster number is set to 6 and 10, respectively. Table \ref{tab:ensemb} shows C-index values using single and ensemble models on MCO-1M and NLST dataset. 
The average c-index across five folds is 0.606 for MCO single and 0.600 for MCO ensemble, respectively. On NLST dataset, the averaged c-index of single model is 0.696 and the ensemble model is 0.695. From the table, it can be seen that ensemble models cannot provide additional power for predictions.

\begin{table}[!htb]\caption{ Results of single and ensemble models}
	\begin{center}		
		\begin{tabular}{ c|cccccc}\hline
			{} & fold 1 & fold 2 & fold 3 & fold 4 & fold 5   \\ \hline
			MCO single & 0.666 & 0.571 & 0.591 & 0.565 & 0.636   \\
			MCO ensemble & 0.665 & 0.564 & 0.585 & 0.549 & 0.637    \\
			NLST single & 0.750 & 0.775 & 0.613 & 0.663 & 0.680    \\
			NLST ensemble & 0.766 & 0.797 & 0.555 & 0.656 & 0.699    \\
			\hline
		\end{tabular}
	\end{center}
	\label{tab:ensemb}
\end{table}

\section{Conclusion}
In this paper, we proposed a deep multiple instance model to directly learn survival patterns from gigapixel images without annotations which make it more easily applicable in large scale cancer dataset. Compared to existing image-based survival models, the developed framework can handle various numbers and sizes whole slide images among different patients. It can learn holistic information of the patient using bag representations and achieve much better performance compared to the ROI patch based methods. Moreover, the flexible and interpretable attention-based MIL pooling can overcome drawbacks from fixed aggregation techniques in state-of-the-art survival learning models. We showed that our approach provides an interpretation of the clinical outcome prediction by presenting reasonable ROIs which is very important in such practical application. Additionally,
We illustrated the proposed method can provide personalized treatment for patients and can be used by doctors to guide their treatment decisions for improving patient lifespan. With future research and development, the proposed approach has the potential to be applied in other tumor types.


\section*{Acknowledgements}
This work was partially supported by US National Science Foundation IIS-1718853, the CAREER grant IIS-1553687 and Cancer Prevention and Research Institute of Texas (CPRIT) award (RP190107).

The authors would like to thank the National Cancer Institute for access to NCI’s data collected by the National Lung
Screening Trial. The statements contained herein are solely of the authors and do not
represent or imply concurrence or endorsement by NCI.

\bibliographystyle{unsrt}  
\bibliography{refs}

\end{document}